\documentclass{cs21proc}

\newcommand\gaia{\textit{Gaia}\,}
\newcommand{\NOBJ}{541\,}   
\newcommand{\NSYS}{336\,}    
\newcommand{\NSTARS}{371\,}  

\newcommand{\NBDS}{85\,}      
\newcommand{\NPLANETS}{85\,}  

\newcommand{\NWD}{21\,}       

\newcommand{\grp}{{$G_\mathrm{RP}$}\,}

\editors{A. S. Brun, J. Bouvier, P. Petit}
\publisher{Zenodo}
\conference{The 21th Cambridge Workshop on Cool Stars, Stellar Systems, and the Sun}
\conferencedate{2022}

\title{The 10 parsec sample in the {\it Gaia} era: first update}
\author{C\'eline Reyl\'e,$^{1}$
        Kevin Jardine,$^{2}$
        Pascal Fouqu\'e,$^{3}$
        Jos\'e A. Caballero,$^{4}$  
        Richard L. Smart,$^{5}$ 
        Alessandro Sozzetti$^{5}$                     }

\affiliation{$^{1}$ Universit\'e de Franche-Comt\'e, Institut UTINAM, CNRS UMR6213, OSU THETA Franche-Comt\'e-Bourgogne, Observatoire de Besan\c con, BP 1615, 25010 Besan\c con Cedex, France \\
			 $^{2}$ Radagast Solutions, Simon Vestdijkpad 24, 2321 WD Leiden, Netherlands\\
			 $^{3}$ IRAP, Universit\'e de Toulouse, CNRS, 14 av. E. Belin, 31400 Toulouse, France\\
			 $^{4}$ Centro de Astrobiolog\'ia (CSIC-INTA), ESAC Campus, Camino bajo del castillo s/n, 28692 Villanueva de la Ca\~nada, Madrid, Spain\\
			 $^{5}$  INAF - Osservatorio Astrofisico di Torino, via Osservatorio 20, 10025 Pino Torinese (TO), Italy 
			 }

\shorttitle{The 10 pc sample}
\shortauthors{C\'eline Reyl\'e et al}

\abs{ 
The nearest stars provide a fundamental constraint for our understanding of stellar physics and the Galaxy. The nearby sample serves as an anchor where all objects can be studied and understood with precise data. This work is an update of the 10\,pc sample published by \cite{2021A&A...650A.201R} that used the unprecedented high precision parallax measurements from the early third data release of the astrometric space mission \gaia. We review this census, all updates being related to close binaries, brown dwarfs and exoplanets.  We provide a new catalogue of \NOBJ\,stars, brown dwarfs, and exoplanets in \NSYS\,systems within 10\,pc from the Sun. This list is as volume-complete as possible from current knowledge and it provides a list of benchmark stars. We also explore the new products made available in the most recent third \gaia data release.}

\begin{document}

\maketitle

\section{Introduction}
Making a census of the nearest stars has been a long term goal in astronomy starting in the end of the nineteenth century, when the first stellar distances were measured from their trigonometric parallaxes. The nearby sample is a fundamental census where all objects can be studied with accuracy. It provides core constraints to understand stellar and galactic physics, as well as ideal targets for exoplanet searches.
Following in the steps of Louise~F. Jenkins, who published a list of 127 stars with their known companions and gathered the knowledge at that time on the neighbours whose distance is less than 10\,pc from the Sun \citep{1937AJ.....46...95J}, we give here the current snapshot of the nearby sample within 10\,pc.

In \cite{2021A&A...650A.201R}, we used the unprecedented high precision parallaxes of \gaia Early Data Release 3 \citep[\gaia\,EDR3,][]{2021A&A...649A...1G} to review the census of objects within 10\,pc. Our first compilation focused on objects observable by \gaia, as a quality assurance test for the 100\,pc Gaia Catalogue of Nearby Stars \citep[GCNS, ][]{2021A&A...649A...6G}. 
 We complemented it with objects not in the \gaia\,EDR3 to get a full 10 pc census, including bright stars, close binaries, brown dwarfs, and exoplanets.
In this paper, we present an update of the 10\,pc census, and explore the new data products (astrophysical parameters, variability, binarity) offered by the third \gaia data release \citep[\gaia\,DR3,][]{2022arXiv220800211G}. 
This new list\footnote{Only available in electronic form
at https://cdsarc.cds.unistra.fr/viz-bin/cat/J/A+A/650/A201, https://gruze.org/10pc/, https://dc.zah.uni-heidelberg.de/10pcsample/q/cone/info, and https://gucds.inaf.it/} provides their astrometry (positions, parallax, proper motions), photometry, radial velocities, and spectral types, when available for \NSTARS stars, \NBDS brown dwarfs, and \NPLANETS exoplanets.

\section{Updates}

\subsection{Added objects}

As predicted by \cite{2021A&A...650A.201R}, the expected additions are cool objects hiding near the Milky Way plane, very close companions resolved by spectroscopic, adaptive optics, or interferometric observations \citep[e.g.][]{2022AAS...24020510V}, and exoplanets that numerically should outnumber the other objects within 10\,pc. Two brown dwarfs and eight exoplanets orbiting M-dwarfs are added to our list.

\begin{itemize}
\item CWISEP J225628.97+400227.3, a Y dwarf that we previously missed from \cite{2021ApJS..253....7K};
\item CWISEP J181006.00--101001.1, a peculiar, metal-poor brown dwarf close to the galactic plane discovered by \cite{2020ApJ...898...77S} and whose parallax was later 
derived by \cite{2022A&A...663A..84L};
\item a companion to GJ~666~B, found from the astrometric orbital solution of \gaia DR3 (see Section~\ref{sec_nss}). This close binarity was wrongly attributed to GJ 666 A by \cite{2010ApJS..190....1R} and \cite{2015MNRAS.453.1439J};
\item GJ 411 c, a long-period planet discovered by \cite{2021ApJS..255....8R} and confirmed by \cite{2022AJ....163..218H};
\item LTT 1445 A c, a second planet transiting the primary star of the triple system LTT 1445 \citep{2022AJ....163..168W};
\item GJ 393 b, a terrestrial planet discovered by \cite{2021A&A...650A.188A};
\item GJ 514 b, a super-Earth planet on an eccentric orbit \citep{2022A&A...666A.187D};
\item GJ 367\,b, a dense, ultra-short period sub-Earth planet \citep{2021Sci...374.1271L};
\item HD 260655\,b and HD 260655\,c, two rocky planets transiting the furthest star of our list \citep{2022A&A...664A.199L};
\item Wolf 1069\,b, an Earth-mass planet in the habitable zone \citep{2023arXiv230102477K}.
\end{itemize}

\subsection{Rejected objects}

Nine low-mass stars and brown dwarfs are removed because they have a better parallax measurement that places them outside of the 10\,pc limit, or because they were wrongly considered as binary objects. 

\begin{itemize}

\item 2MASS J16471580+5632057 has a more accurate parallax of 42.9 $\pm$ 2.1\,mas by \cite{2020AJ....159..257B};

\item 2MASS J07584037+3247245 has a better \gaia EDR3 parallax of 91.92 $\pm$ 1.73\,mas than the one that we considered \citep{2021ApJS..253....7K};

\item CFBDS J213926+022023 A and B is actually a single object \citep{2021ApJS..253....7K} and has a better parallax of 96.5 $\pm$ 1.1\,mas by \cite{2021ApJ...911....7Z};

\item GJ~666\,Ab, a close companion wrongly attributed to GJ~666\,A instead of GJ~666\,B by \cite{2010ApJS..190....1R} and \cite{2015MNRAS.453.1439J};

\item GJ 748 AB has a more accurate and robust {\em Hubble} parallax of 98.4 $\pm$ 0.3\,mas taking into acount the effects of binarity \citep{2016AJ....152..141B};

\item GJ 424 B \citep{2006AJ....132..994D} was not confirmed with adaptive optics \citep{2015MNRAS.449.2618W} and long term high precision radial velocity monitoring rules out the proposed companion candidate  \citep{2017AJ....153..208B};

\item UPM J0815--2344 B is a background object that is close to UPM J0815--2344 and was wrongly attributed as a physical companion by \cite{2018AJ....155..265H};

\item WISE J081117.81--805141.3 was wrongly listed in the 10\,pc sample. Having a parallax of 98.5 $\pm$ 7.7\,mas \citep{2014ApJ...796...39T}, it is now part of the candidate list.
\end{itemize}

\subsection{Candidates}

Star and brown dwarf candidates are tabulated in the list (numbered from {\tt NB\_OBJ} equal 1001 to 1021). They are mostly brown dwarfs that have large parallax uncertainties and are still compatible with a parallax larger than 100\,mas at the $1\sigma$ level.\\

Exoplanet candidates are given in the {\tt COMMENT} field of the list. Three new candidates are GJ~411\,d \citep{2022AJ....163..218H}, LTT~1445~A\,d \citep{2022arXiv221009713L}, and Proxima Cen\,d \citep{2022A&A...658A.115F}. None of our previous candidates have been confirmed yet. However, we flagged the planet around Barnard's Star announced by \cite{2018Natur.563..365R} as controversial rather than candidate based on the stellar activity study by \cite{2021AJ....162...61L}.\\ 

We also add a note on GJ 229 B: the high dynamical mass derived by \cite{2021AJ....162..301B} may denote the existence of an unseen companion.\\


\subsection{Additional objects with a \gaia parallax larger than 100\,mas}

Three additional objects have a \gaia EDR3 parallax larger than 100\,mas. The random forest classification procedure used for the construction of the GCNS \citep{2021A&A...649A...6G} found that these objects have a bad solution based on astrometric quality assurance parameters. Their probabililty of having a good astrometric solution is very low (0.076, 0.082, 0.013), much lower than the 0.38 threshold defined as the reliable astrometry probability.
Thus they were rejected by \cite{2021A&A...650A.201R} without further discussion although we noticed inconsistencies with WISE and PanSTARRS data as described below. 

Using other selection criteria, the Fifth Catalogue of Nearby Stars (CNS5) recently published by \cite{2022arXiv221101449G} kept two of 
the three objects with \gaia EDR3 parallax larger than 100\,mas.
We believe that they deserve more discussion,
because the
three sources, listed below, are blended and have rather low total proper motions (13, 29, 69 mas\,a$^{-1}$) compared to the usual values of the nearby sample (mean, minimum, maximum, and standard deviation, of 1319, 68, 10393, and 1246 mas\,a$^{-1}$, respectively).

\begin{itemize}
\item  The faint $G$ magnitude (20.6\,mag) of \gaia EDR3 4318384355378007424 together with being nearby (\gaia EDR3 parallax of 101.08 $\pm$ 3.47\,mas) would imply a red source but it appears blue in PanSTARRS. It has no \grp magnitude whereas as a nearby red object it should be detected. A very close (1\,arcsec) bright source probably makes the \gaia observation difficult. It also was also rejected during the selection process of CNS5;

\item \gaia EDR3 6305165514134625024 has a \gaia EDR3 parallax of 174.02 $\pm$ 1.90\,mas. A $G=$ 20.4\,mag at this close distance is supposed to be a mid-T dwarf, but its $W1$ = 16.9\,mag is about 4.5\,mag fainter than a nearby mid-T should be. Looking at the PanSTARRS and WISE images, we instead interpret it as a red background object blended with a blue object. It is considered as a new addition to the 10\,pc sample in CNS5. For the moment we tabulate it as a candidate ({\tt NB\_OBJ} = 1021) with a low probability to be an exotic object;

\item \gaia EDR3 4479498508613790464 is also considered as a new addition to the 10\,pc sample in CNS5. The \gaia EDR3 parallax of 121.98 $\pm$ 0.94\,mas is probably wrong based on recent spectroscopic observations  by Kirkpatrick et al (in prep), who attributed an M2\,V spectral type that matches the observed colours.

\end{itemize}

\begin{figure*}[h]
	\centering
	\includegraphics[width=0.45\linewidth]{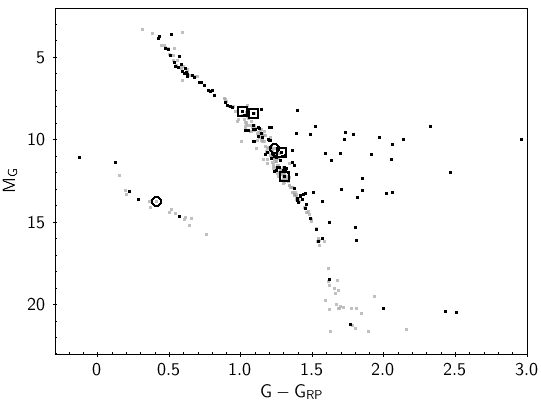}
	\includegraphics[width=0.45\linewidth]{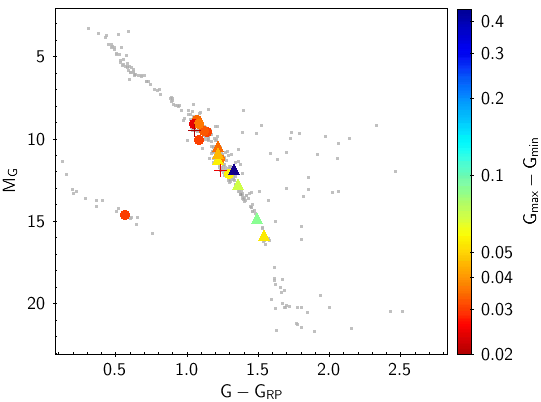}
	\caption{Colour absolute magnitude diagram of the 10\,pc sample with \gaia photometry (352 objects). Left panel: The objects in binary (or higher multiplicity) systems and unresolved binaries are highlighted in black. The objects found in \gaia~DR3 non-single stars tables are shown with open symbols, where the two circles are for a planet or candidate planet. Right panel: The stars with variability parameters in \gaia~DR3 are shown with circles (solar-like) and triangles (short-timescale). The crosses are part of the \gaia Andromeda Photometric Survey (see text). The colour bar gives the maximum value of the variation.}
	\label{fig:fig_nss}
\end{figure*}
 \section{The 10\,pc sample in \gaia DR3 products}

The \gaia third data release is the outcome of the processing of data collected during the first 34 months of the mission. It was been done in two steps. The Early Data Release 3 (EDR3), published on 3 December 2020, provides new astrometry and photometry of the sources with radial velocities from the second \gaia data release. The Data Release 3 (DR3), published on 13 June 2022, provides more radial velocities and a wealth of new data products, such as non-single stars, variability properties, and astrophysical parameters. In what follows, we show what kind of information can be found in \gaia DR3 for the 10\,pc sample.

\subsection{Non-single stars and exoplanets}
\label{sec_nss}

Six objects, five M dwarfs and one white dwarf, are in the table \texttt{gaiadr3.nss\_two\_body\_orbit}, which contains orbital models 
compatible with an orbital two-body solution. A selection of parameters, such as period, periastron argument, eccentricity or inclination, is provided in the table, depending on the solution type \citep[namely: astrometric, spectroscopic, photometric; see][]{2022arXiv220605595G}. For objects with an astrometric solution, new values of the parallax and proper motions, taking into account the orbital motion, are provided. We updated these values, which are more accurate than those given in the main catalogue, in our 10\,pc sample. In addition, the table \texttt{gaiadr3.binary\_masses} provides an estimate of the masses and flux ratios, or lower and upper limits of them. We give below details on the parameters for the six objects with a non-single star solution. They are shown with open symbols in the colour absolute diagram in Fig.~\ref{fig:fig_nss}, left panel.

\begin{itemize}
\item GJ 1230 AC has an orbital solution from spectroscopy (SB2) and thus the secondary mass can be estimated. The value is 0.299\,M$_\odot$, confirming the low-mass star type of GJ~1230~C with no spectral type. The estimated orbital period is 2.53\,days;
\item GJ 867 AC received independent, astrometric and spectroscopic, orbits. The derived secondary mass, 0.635\,M$_\odot$, confirms the low-mass star nature of GJ~867~C with no spectral type. The estimated orbital period is 4.08\,days;
\item Wolf 227 AB has an orbital solution from astrometry. The secondary mass ranges from 0.046 to 0.364\,M$_\odot$. The lower value is compatible with the statement that it may be a brown dwarf from its mass estimate from \cite{2018AJ....155..125W}. The period is found to be 10.59\,days;
\item GJ 666 B has an unseen companion detected from its astrometric orbit. The secondary mass ranges from 0.169 to 0.734\,M$_\odot$, and the period is 87.91\,days;
\item GJ 876 b is the only planet at less than 10\,pc detectable by \gaia according to \citet{2021A&A...650A.201R} that was detected in DR3.
Its true mass derived from the astrometric orbit is 3.6\,M$_\mathrm{Jup}$, larger than any of the various estimates in the literature (which range between 2.0 and 2.7\,M$_\mathrm{Jup}$). In \citet{2022arXiv220605595G} some discussion is provided on the possible nature of the discrepancy;
\item L 88-59, a white dwarf, has an astrometric orbital solution with a secondary mass ranging from 0.007 to 0.838\,M$_\odot$ and a period of 33.65\,days. The lower mass value of the secondary makes it a planet candidate, which is listed in the dedicated list of \gaia exoplanets maintained at {\tt https://cosmos.esa.int/web/gaia/ exoplanets}.
\end{itemize}

\subsection{Variable stars}

Up to 19 stars in our sample are found in the variability tables (\texttt{gaiadr3.vari\_summary, gaiadr3.vari\_classifier\_result, gaiadr3.vari\_short\_timescale, gaiadr3.vari\_rotation\_modulation}), which give various parameters derived from multi-epoch observations \citep{2022arXiv220606416E}. They are shown with coloured symbols in Fig.~\ref{fig:fig_nss}, right panel. Nine are solar-like variable stars, indicating a variable phenomena similar to those observed in the Sun, mainly due to the evolution of its magnetic active regions (dark spots and bright faculae unevenly distributed over the stellar surface). They are all M dwarfs (GJ~ 625, AN~Sex, GJ~1151, L~49-19, G~19-7, MCC~135, BD+43~2796, BD+16~2708A), except for one white dwarf (HD~100623~B). Another seven stars, all M dwarfs, are short-term candidates with timescales between a few tens of minutes to one day (Ross~248, DENIS~J104814.6-395606, GJ~643, GJ~486, L~173-19, LP~655-48, BD+61~195B). G~19-7 and GJ~867~B have in addition rotational modulation and, therefore, their stellar rotation period can be determined from the analysis of their light curve (1.20\,days and 1.99\,days, respectively). Finally, GJ~15~A and GJ~15~B are also part of the variability table because 
photometric data were obtained as part of the \gaia Andromeda Photometric Survey \citep{2022arXiv220605591E}. 

\subsection{Radial velocities and kinematics}

\begin{figure}
	\centering
	\includegraphics[width=0.95\linewidth]{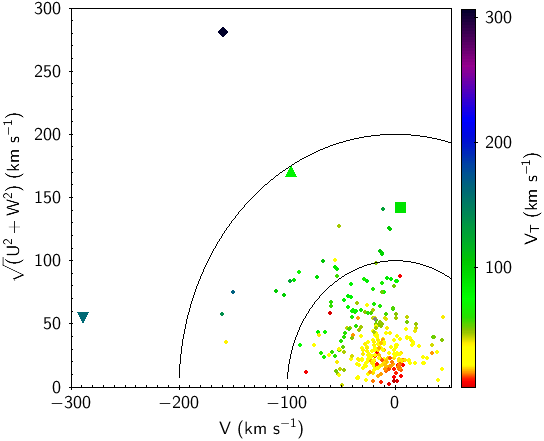}
	\caption{Toomre diagram for the 309 objects with a radial velocity measurement, coloured by their transverse velocity. The circles with total velocity of 100 and 200 km\,s$^{-1}$ are indicative values to delineate thin-disc, thick-disc, and halo stars. Square: Barnard's star, diamond: HD\,103095, triangle down: Kapteyn's star, triangle up: 2MASSW J1515008+484742.}
	\label{fig:fig_kine}
\end{figure}

All stars and brown dwarfs in the sample have measurements of parallax and proper motion, which allows us to compute their transverse velocity, $V_T$.
In addition, \gaia DR3 provides new radial velocities for 23 stars, and more accurate radial velocities than previously measured for another 24 stars \citep{2022arXiv220605902K}, leading to 309 stars with full kinematics in the 10\,pc sample. We exclude the erroneous large value of $-414$\,km\,s$^{-1}$ for the white dwarf EGGR\,290, due to the lack of white dwarf templates in the radial velocity determination \gaia pipeline. For those stars we are able to compute the local standard of rest-corrected space velocities in the Galactic reference frame ($U, V, W$).  Figure~\ref{fig:fig_kine} shows the resulting Toomre diagram 

Most of the nearby sample lie in the thin-disc region. There are, however, a few remarkable exceptions. The star with the highest tangential velocity (square) is Barnard's star, which has a prograde motion and lies in the thick-disc region. 2MASSW J1515008+484742 (triangle up) is close to the thick-disc-to-halo boundary. Kapteyn's star (triangle down) and HD\,103095 (diamond) have the largest total velocities and lie in the halo region. The latter also has the largest secular accelaration, with a value of $10.510$\,m\,s$^{-1}$\,a$^{-1}$, much larger than the value of $6.755$\,m\,s$^{-1}$\,a$^{-1}$ for Kapteyn's star. HD\,103095 is also shown in a chemistry plane in Sect.~\ref{sec:ap}.

\subsection{Astrophysical parameters}
\label{sec:ap}

\begin{figure}
	\centering
	\includegraphics[width=0.95\linewidth]{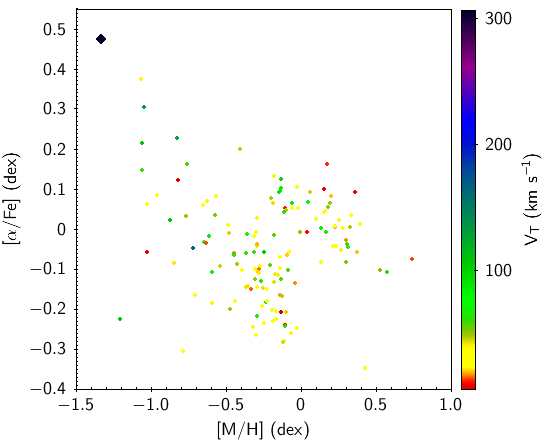}
	\caption{[$\alpha$/Fe] vs. [M/H] diagram for a subsample of 248 stars with good quality parameter measurements in \gaia\,DR3, coloured by their transverse velocity. The diamond shows HD\,103095.}
	\label{fig:fig_chim}
\end{figure}

Several astrophysical parameters have been derived from \gaia photometry and spectroscopy and, therefore, part of our sample has a determination of the effective temperature, gravity, metallicity and, in a few cases, global abundance of $\alpha$-elements with respect to iron \citep{2022arXiv220606138A,2022arXiv220605541R}.  
Furthermore, some sample stars also have mass, luminosity, radius and age determinations from \gaia\,DR3 \citep{2022arXiv220605870G}.
We did not attempt to add these values in our list, since it would ask for a large work to get a consistent picture with other values found in the literature \citep[e.g.][]{2020A&A...642A.115C,2020MNRAS.492.5844R,2021A&A...656A.162M,2022MNRAS.516.3802C}. 
As an illustration, we show in Fig.~\ref{fig:fig_chim} the [$\alpha$/Fe] vs [M/H] plane. This plane is often used to investigate stellar populations, \citep[see e.g. earlier works by ][]{1998A&A...338..161F,2006MNRAS.367.1329R,2013A&A...554A..44A}, who showed that the disc of the Milky Way is composed of $\alpha$-rich and $\alpha$-poor stars. We only plot the stars with the 13 parameter quality flags $\leq 1$ \citep[see Table\,2 in][]{ 2022arXiv220605541R}. We also applied the calibrations on [M/H] and [$\alpha$/Fe] as a function of their log\,g as defined by \cite{2022arXiv220605541R}. In Fig.~\ref{fig:fig_chim}, HD\,103095 (diamond) lies in the upper ($\alpha$-rich) left (metal-poor) part of the diagram, pointing to an old population star, compatible with its extreme kinematics (see Fig.~\ref{fig:fig_kine}).

\section{Conclusion}
We provide an update of the catalogue of all objects closer than 10\,pc from the Sun. This list shows the high variety of objects contained in the immediate vicinity of the Sun.
It contains \NOBJ\ objects divided between \NSTARS\ stars, including \NWD\ white dwarfs, \NBDS\ brown dwarfs, and \NPLANETS\ confirmed exoplanets in \NSYS\ systems.
It contains the most recent astrometry from the last \gaia data release when available. As \citep{2021A&A...650A.201R} already pointed out, the updates concern close binaries, brown dwarfs, and exoplanets, and we expect that in the future the number of stars and brown dwarfs will be superseded by exoplanets. In addition, we explore the new products offered by the most recent \gaia DR3, including astrophysical parameters, additional radial velocities, non-single star orbital solutions, and variability parameters. This list provides a set of benchmark stars to be studied in detail with current and forthcoming instruments. More parameters, in particular on the non-single stars (including exoplanets) are expected in the forthcoming \gaia data releases.


\section*{Acknowledgments}
{The authors thank Kevin App, Fr\'ed\'eric Arenou, Laurent Eyer, Brian\,D. Mason, and Andrei Tokovinin for fruitful exchanges.
CR acknowledges financial support from the "Programme National de Physique Stellaire" (PNPS) of CNRS/INSU, France.
}

\bibliographystyle{cs21proc}
\bibliography{cs21reyle.bib}

\end{document}